\documentclass{aastex}

\begin{document}

\title{RELATIVISTIC CONIC BEAMS AND SPATIAL DISTRIBUTION OF GAMMA-RAY BURSTS}

\author{Heon-Young Chang and Insu Yi}

\affil{Korea Institute For Advanced Study, Seoul, 130-012, Korea}
\email{hyc@kias.re.kr, iyi@kias.re.kr}

\begin{abstract}

We study the statistics of gamma-ray bursts, assuming that gamma-ray bursts are 
cosmological and they are beamed in the form of a conical jet with a large 
bulk Lorentz factor $\sim 100$. In such a conic beam, the relativistic ejecta 
may have a spatial variation in the bulk Lorentz factor and the density 
distribution of gamma-ray emitting jet material. 
An apparent luminosity function arises because the axis of the cone is randomly
oriented with respect to the observer's line of sight. The width and the shape 
of the luminosity function are determined by the ratio of the beam opening 
angle of the conical jet to the inverse of the bulk Lorentz factor, 
when the bulk Lorentz factor and the jet material density is uniform on the
photon emitting jet surface.
We calculate effects of spatial variation of the Lorentz factor and the spatial
density fluctuations within the cone on the luminosity function and the 
statistics 
of gamma-ray bursts. In particular, we focus on the redshift distribution of
the observed gamma-ray bursts. The maximum distance to
and the average redshift of the gamma-ray bursts are strongly affected by 
the beaming-induced luminosity function. 
The bursts with the angle-dependent Lorentz factor which peaks at the center 
of the cone have substantially higher average gamma-ray burst redshifts. 
When both the jet material density and the Lorentz factor are inhomogeneous
in the conical beam, the average redshift of the bursts could be 5 times higher
than that of the case in which relativistic jet is completely homogeneous 
and structureless. Even the simplest models for the gamma-ray burst jets and
their apparent luminosity distributions have a significant effect on the
redshift distribution of the gamma-ray bursts.

\end{abstract}

\keywords{gamma rays:bursts}

\section{INTRODUCTION}

The BATSE experiment on the {\it Compton Gamma Ray  Observatory} and the study
of the afterglows (e.g., Piran 1999 and references therein).
have established that the gamma-ray bursts (GRBs) are 
cosmological \citep{mao92, meegan92,pi92}. 
Even though the distance scale seems settled  \citep{met97}, 
it appears that uncertainties remain in the total energy and 
the burst rate of GRBs \citep{ku99, kupi99}. 
These two important issues depend on the level of beaming in GRB emission.
That is, the issues critically depend on whether the geometry of the gamma-ray 
emitting ejecta is spherical or jet-like  \citep{har99, kul99, mes99, sari99}. 
A number of authors have studied energetics and geometry 
of the ejecta \citep{maoyi94, rho97, pan98, rho99, mod00}.
It is also important to put constraints on the width of the luminosity
function by comparing the observed intensity distribution with those predicted
by a physical model \citep{maoyi94, yi94}. In essence, the rate, the energy, 
and the luminosity function of GRBs are all closely related to 
whether or not the geometry  of the ejecta is spherical.

Two most frequently quoted statistics in GRB observations are $<V/V_{max}>$ 
and $\log N (>F) - \log F$, 
where $F$ refers to the peak flux (or peak count rate) 
and $N$ denotes the number of GRBs with fluxes higher than $F$ (e.g., Yi 1994). 
These two quantities contain information on the lumonosity function of GRBs 
and the spatial number density of the sources. 
A value of $<V/V_{max}>$ consistent with that of an observed sample is
a necessary condition but not a sufficient condition for a luminosity 
function $\Phi(L)$ which is neither directly observable nor theoretically
well undertood. The luminosity function of GRBs can 
be obtained for an assumed source distribution $n(z)$ such that 
the calculated $\log N(>F) - \log F$ fits the observed distribution, 
and vice versa. 
The density $n(z)$ refers to the rate of GRBs per
unit time per unit comoving cosmological volume. 
However, due to the very nature of  $N(>F)$, 
which is the convolution of $n(z)$ and $\Phi(L)$, 
one almost always obtains  
$n(z)$ for a given  $\Phi(L)$ such that the theoretical  $\log N(>F) - \log F$ 
curve fits the observed intensity distribution. Therefore, in order to extract 
information concerning  $n(z)$ or $\Phi(L)$, one has to assume one of these 
two functions or to develop a techinique to separate the effects of these 
two unknown functions 
(Horack and Emslie 1994; Horack et al. 1994; Ulmer et al. 1995; 
Ulmer and Wijers 1995). 
It is therefore of great interest to construct $\Phi(L)$ on the basis of 
a physical model, which is one of our major goals in this {\it Letter}.
Since there remain uncertainties in GRB engine models, we focus on the
consequences of the conical beaming without specifying how a beam is
formed in a physical engine model.

Using the first BATSE catalog of gamma-ray bursts \citep{fishman94},
\citet{maoyi94}  
studied the effects of the relativistic bulk motion in a conical beam on the 
statistics of gamma-ray bursts. They found that the luminosity function is 
naturally introduced by the random distribution of the space orientation of 
the cone axis and that the case of the standard candle is not easily
distinguished from that of the beaming-induced luminosity function with a
sharp peak.
This is especially the case for large beam opening angle and the large Lorentz
factor $\gamma$, as one may expect. 
Different Lorentz factors and opening angles however result in 
non-trivial changes for the distances to GRBs and especially the highest
redshift of or the maximum distance to the most distant GRB for a given sample.
For instance, the maximum redshift $z_{max}$ 
increases as the ratio of the opening angle to  $\gamma^{-1}$ decreases.
We modify the conical beam model by allowing a spatial variation of $\gamma$ 
and the density profile of gamma-ray emitting electrons on the photon-emitting  surface of the cone.

From numerical simulations of relativistic jets 
\citep{marti97, renaud98, rosen99} and observations of the 
astrophysical jets \citep{zen97, spruit00},
it is clear that jets do have some significant structure in them and the bulk
Lorentz factors evolve as the jets propagate. Therefore, it is plausible
to extend the simplest jet model such as that of \citet{maoyi94}.
In a more realistic jet model, the bulk Lorentz factor has a spatial profile 
at the surface where the observed gamma-ray emission occurs and the spatial 
electron density distribution is significantly inhomogeneous. 

In $\S$ 2 we begin with a  brief presention of data we use, 
which are parts of the BATSE 4B catalog  \citep{paciesas99}, and 
 we describe our conical jet geometry, 
following \citet{maoyi94}, in $\S$ 3 we present results. 
Finally, we conclude with summary of our results and discussions in $\S$ 4.

\section{OBSERVATIONAL DATA AND MODEL}

The BATSE 4B catalog  \citep{paciesas99} provides 1637 triggered GRBs
detected from 1991 April through 1996 August. We use the bursts which are 
detected on the 1024 ms trigger time scale. We choose the bursts of which peak 
count rates are above 0.4 ${\rm photons \hspace{1mm} cm^{-2} s^{-1}}$ in order
to avoid the threshold effects (cf. Mao and Yi 1994). 
Of those bursts, we select the GRBs whose ${\rm C_{max}/C_{min}}$ is greater
than $ 1.0$, which gives a sample of 651 bursts.
The energy range in which the peak flux is measured is $50 - 300 \hspace{1mm}
 {\rm keV}$.  The peak fluxes of bursts vary by about 2 orders 
of magnitude in this data  set.

Throughout this {\it Letter} we adopt  the simplest cosmological model; 
the universe is flat with the density parameter $\Omega_{0}=1$, 
the Hubble constant is 65 ${\rm kms^{-1}Mpc^{-1}}$, the cosmological constant
is absent, all the GRBs are 'standard bursts' with an identical power-law  
spectrum, and the rate of bursts per unit comoving volume per unit comoving 
time is constant (cf. Mao and Paczy${\rm \acute{n}}$ski 1992; Yi 1994).

In our beaming model, the ejecta is flowing outward relativistically 
in a cone with the geometrical opening angle $\Delta \theta$. 
The observed gamma-ray emission is produced at radius $R$ from the central
engine where the radiation 
first becomes optically thin. In the minimal beaming model of \citet{maoyi94},
the ejected material has the same bulk Lorentz factor $\gamma$ at this distance 
and the photon-emitting electrons' density at the surface of the cone is
uniform. In this {\it Letter} those simplifications are relaxed as described 
below. 

In the cylindrical symmetry one may consider the colatitude alone, which is
defined by the angle between the line of sight and the symmetry axis of the 
cone. Besides the spherical coordinate system centered on the central engine,
we introduce an auxiliary spherical coordinate system with the $z^{'}$-axis 
along the symmetry axis of the cone. It can be  shown that the angle $\Theta$
 between a position of a direction within the cone and the  line of sight is
given by
\begin{equation}
\cos\Theta=\cos\theta^{'}\cos\theta-\sin\theta^{'}\sin\phi^{'}\sin\theta.
\end{equation}

The monochromatic flux received by a local observer  at a distance $D$ from 
the  source, taking the cosmological redshift effects into account, reads
\begin{eqnarray}
F(\nu,\theta)&=&\frac{(1+z)R^{2}}{D^{2}_L(z)} 
\times \int^{2\pi}_{0}d\phi^{'}\nonumber\\
&& \times \int^{\Delta \theta}_{0} \sin \theta^{'} d \theta^{'} \cos \Theta
\Gamma^{3} I_{0}[\nu(1+z)\Gamma^{-1}],
\end{eqnarray}
where $D_L(z)=2c/H_0[1+z-(1+z)^{1/2}]$, $c$ and $H_0$ being the speed of light 
and the Hubble constant respectively, and 
$\Gamma =[\gamma(1-\beta \cos \Theta)]^{-1}$, $\beta = (1-\gamma^{-2})^{1/2}$. 
And  we have used the   relation $\nu =\Gamma \nu_{0}$; here $\nu_{0}$ is the
  corresponding frequency in the comoving frame. In this model we have ignored
the structure of light curves due to the relative time delay of radiation from
different parts of the cone, and other complicated cosmological effects.

We define the terms, excluding terms for the cosmological information 
in the above equation (2), as the local peak count rate,
\begin{eqnarray}
P_{loc}(\theta)&=&\int \nu^{1-\alpha} d\nu \times  R^{2}\int^{2\pi}_{0}d\phi^{'}\nonumber\\
&& \times \int^{\Delta \theta}_{0}\sin \theta' \cos \Theta \Gamma^{2+\alpha} 
d \theta'.
\end{eqnarray}
The maximum local peak count rate is achieved at $\theta=0$ and can be obtained 
analytically : 
\begin{eqnarray}
P_{loc,max}(\theta=0)&=&\int \nu^{1-\alpha}
 d \nu \frac{2 \pi R^{2}}{\beta^2 \gamma^{2+\alpha}}  \nonumber\\
&& \times [\frac{1}{\alpha}(x^{\alpha}_u-x^{\alpha}_l) -
\frac{1}{1+\alpha}(x^{1+\alpha}_u-x^{1+\alpha}_l)],
\end{eqnarray}
where $x^{-1}_u=1-\beta \cos \Delta \theta$, and  $x^{-1}_l=1-\beta$.

Based on the randomness of  the direction of the cone axis with respect to the 
line of sight, we obtain the probability function which directly reflects 
the angle between the line of  sight and the direction of  the cone,  $\theta$. 
Assuming that the cone is uniformly distributed in space, the probability 
function for a one-sided cone is given by 
\begin{equation}
p(\theta) d \theta = \frac{1}{2} \sin \theta d \theta,
\end{equation}
where $\theta$ is between $0$ and  $\pi$. Since the local peak count rate is 
a function of the orientation of the cone, one may translate  the probability
function of the cone's angle into the probability
function of the local peak count rate. 

In order to obtain the luminosity function and compare with observations, 
we calculate the local peak count rate in a fixed frequency range. 
We allow variations of $\gamma$ and the electron density at the surface of the 
cone by introducing a window function for $\gamma$ and electron density, $N_e$.
The window function is axisymmetric with respect to the symmetry axis 
of the cone. The window function we adopt is the Gaussian function centered 
at the center of the cone:
\begin{eqnarray}
W(\theta')=\exp [-A(\frac{\theta^{'}}{\Delta \theta})^{2}],
\end{eqnarray}
where $A$ is a constant, 
$\Delta \theta$ is a given opening angle, and $\theta^{'}$
varies from $0$ to $\Delta \theta$.

\section{RESULTS}

The probability function of the local peak count rate is shown in  
Figure 1 
for three different ratios of $\Delta \theta$ to $\gamma^{-1}$, 
in the case where $\gamma$ is uniform over the photon-emitting surface 
of the cone. For a fixed $\gamma$, as the opening angle 
increases the peak of the probability of the local peak count rate 
becomes higher and narrower, as one may expect. A model with a large value of 
Lorentz factor ($\Delta \theta \gg \gamma^{-1}$) essentially  provides a
luminosity function indistinguishable from that of the standard candle model. 
For a given value of the ratio, the whole distribution function
moves vertically when the absolute value  of the Lorentz factor changes. 
This is because the local peak count rate is decreasing faster with the angle
for larger $\gamma$. In the extreme case
where $ \Delta \theta \ll  \gamma^{-1} $ the distribution is  a power law  with 
the index of $ -1/3$, as expected from the analytical result \citep{yi93}. 

In Figure 2, we show the probability distribution functions for 
both the varying bulk Lorentz factor and the inhomogeneous electron density.
We assume the axisymmetry of the Lorentz factor profile around the cone
axis, $\gamma=\gamma(\theta^{'})$, and hence the Lorentz factor profile
could mimic a simplified model for the jet-environment drag.
At the center of the cone, $\gamma$ has the maximum value and decreases with 
$\theta'$. As $\gamma$ decreases more steeply with $\theta'$, the
$\log p_{L}(\log P_{loc})$ 
shows a smaller peak at $\log (P_{loc}/P_{loc,max})=0$,
and a  higher level of the 'tail' of the  $\log p_{L}(\log P_{loc})$. 
It is  because this type of $\gamma$ effectively reduces the 'average' value 
of $\gamma$ over the  cone and the 'effective opening angle' simultaneously. 
For a very narrow window function it essentially reduces to 
the pencil-beam case.
The photon-emitting electrons are supposed to be distributed according to the
$\gamma$ distribution such that the local electron number density is inversely
proportional to the square of the bulk Lorentz factor, $\gamma$. That is, 
the electron number density is a function of the angular position in the cone. 
In this profile, the electron number density increases outwards from the center 
of the cone, since $\gamma$ is a decreasing function. In effect, the cone is  
reminiscent of the hollow cone.

Once the probability distribution is obtained, we are in a position to 
calculate $\log N(>F) - \log F$ with an assumed spatial distribution of GRBs, 
$n(z)$. In Figure 3, we compare the
cumulative intensity distribution curve produced by our luminosity function
with observational data. Since the goal of this study is looking at effects
of the conic beam with varying distributions of 
$\gamma$ and the electron density 
over the surface of the cone, we assume there is  no source evolution, that is,
$n(z)=n_{0}$. 
We adopt the power law index $\alpha$ of unity instead of 2
 (cf. Mao and Yi, 1994). This is a simplification in a sense that observed
GRBs show various power-law indices. This value, however, represents
the averaged power law indices of observed GRBs \citep{band93,mal96,preece00}.
All the theoretical cumulative probability distribution with the 
luminosity functions shown in Figure 2  are plotted in 
Figure 3
along with the observed distribution. 
The curves shown in Figure 3 are best-fit functions 
determined by the  Kolmogorov-Smirnov  (K-S) test for each parameter set. 
Parameters in our model are not sensitive enough to constrain the 
luminosity function. 

For a given set of model paramters, 
$z_{max}$ plays a role of a parameter in the 
K-S test in that the luminosity function studied here optimizes  
$\log N(>F) - \log F$ to fit observational data. The obtained $z_{max}$'s 
in this procedure
 are shown in the Table 1.
These  $z_{max}$'s obtained with the 4B BATSE catalog are greater than those 
for the 1B BATSE catalog by about $0.1$ on average, which indicates that
we are seeing fainter GRBs in the 4B BATSE than in the 1B BATSE catalog
and therefore more distant GRBs. 

It shows that the redshift of the most distant GRBs
becomes larger as $\gamma$ falls more steeply from the cone center.
As shown in the Table 1, our toy models with the narrow Gaussian profiles
could easily reproduce $z_{max}$ values as high as the highest reported 
$z_{max}=3.42$ (quoted from Bulik (1999), see references therein). This
simple case indicates that the effects of the luminosity function have
significant implications on the cosmological spatial distribution of GRBs.
It is interesting to see that such a high
redshift would be hard to explain in the homogeneous beam model with 
a constant $n(z)$. The effects of the Gaussian window on the
values of $z_{max}$ are substantial enough for further comments.
The model for a narrow window with $A=4$ gives $z_{max}=3.7 
({\rm average \hspace{2mm} redshift} <z> = 1.03)$ while the uniform beam
model (i.e. without any variation of $\gamma$) gives $z_{max}=1.6 (<z> = 0.51)$. 
The average value of redshifts, $<z>$, is taken over 
the cumulative redshift distribution function shown in Figure 4. 
The cumulative redshift distribution is defined by
\begin{eqnarray}
{\cal N}(z^{'})=\frac{N(z^{'}<z<z_{max})}{N(0<z<z_{max})}
\end{eqnarray}
where
\begin{eqnarray}
N(z^{'}<z<z_{max})=\int^{z_{max}}_{z^{'}}\frac{4 \pi}{1+z}n(z)r^2(z)dr(z).
\end{eqnarray}

The cumulative redshift distributions shown in Figure 4 also indicate that
the luminosity distribution induced by the beaming has a significant 
implication on the GRBs distances in the observed samples. When the beam
is sharply peaked at the beam center
 (solid curves in the upper panel in Figure 4)
the fraction of the high redshift GRBs (e.g. $z>3$) could be as high as $\sim
10\%$ while the broad beam case (e.g. $A=1/8$) essentially rules out any
high redshift GRBs.  The cumulative distribution slowly decreases with 
the redshift when the $\gamma$ in the conic beam is decreasing rapidly 
in a sense that GRBs are spead out in a broader region 
beyond the averaged $z$ for the highly concentrated beam.
That is, the ratio of $z_{max}$ to $<z>$ is 3.59 and 3.14 when $A=4$ and $A=1/8$,
respectively. 
The beams' electron density structure also has a
similar effect on the redshifts of GRBs. The effects of the luminosity
function have to be explicitly considered when observed high redshift values 
are interpreted \citep{kth98}.
For instance, in the standard candle case, a significant
probability for high redshift GRBs could directly imply a substantial
source evolution effect. However, the beaming induced luminosity function
could make this simple interpretation much uncertain \citep{blain00}.

\section{DISCUSSION}

The theoretical models for GRBs are abundant. Despite remarkable progresses
in understanding physical mechanisms involved in these models, the GRB
prompt emission mechanisms and engine models have so far been unable to
constrain the extent of beaming and the luminosity distribution of GRBs. This
in turn has been a major uncertainty in interpreting the observed flux data
in terms of the cosmological spatial distribution of the bursts. In this
regard, the present work has shown that the simple beaming models and
their resulting apparent luminosity functions have significant effects
in interpreting the observed data. If the GRBs are indeed standard candles
with a single well-defined luminosity, the spatial distribution of GRBs
in connection with the cosmic star formation rate could be translated into
the cosmological source evolution. However, the luminosity distributions we
have considered affect the maximum redshift and the average redshift significantly.
It is therefore important to derive a theoretical luminosity function for
a given GRB model. 

The jet models we adopted are obviously over-simplified. Despite this major
drawback, the models capture the essential ingredients of the beamed 
relativistic jets concerning the apparent luminosity function. One of the
major uncertainties is that the jets and GRB sources differ greatly and
GRB luminosities and jets' physical conditions are intrinsically different
in each source. Given the wide range of burst durations and the diverse
burst types  (Fishman and Meegan 1995, and references therein),
 such a possibility cannot be ruled out. 
If this is indeed the case, our standard source approach is not applicable.

\acknowledgments
We thank C. Kim, H. Kim, and K. Kwak for useful discussions and especially C. Kim for
her help with Figure 3. IY is supported in part by the KRF grant 
No. 1998-001-D00365.

\clearpage

\figcaption[fig1.ps]{Logarithm of the probability distribution 
$p_{L}(\log P_{loc})$ as a function of $\log P_{loc}$. The local peak count 
rate is scaled  to its maximum value. In this example, $\gamma=10^2$.
>From bottom to top, the opening anlges are $0^{\circ}.1, 1^{\circ}, 3^{\circ}$. 
The photon spectral index $\alpha$ is unity.
Note that for  $\Delta \theta = 0^{\circ}.1$ case, it is basically a power-law
function with the index of $-1/4$. 
\label{fig1}}

\figcaption[fig2.ps]{Plots similar to Fig. 1. In the upper panel, 
values of $\gamma$ varies over the surface of the cone. The window function 
adopted is the Gaussian, $\exp [-A(\frac{\theta^{'}}{\Delta \theta})^{2}]$ 
such that at the center of the cone $\gamma$ has an original constant value, 
$\gamma_{0}$. Results of this study do not depend specific details of the 
window function. In the lower panel, the electron number density also
varies over the surface of the cone. The electron density varies as inversely 
proportional to the square of $\gamma$ in this example.
The thick solid lines correspond to the constant 
$\gamma_{0}$ case with $\alpha =1$, the thin solid lines to $A=2$,
the dotted lines $A = 1$, the short dashed lines $A = 1/8$. 
The opening angle of the cone is $1^{\circ}$ and $\gamma$ is $100$. 
\label{fig2}}

\figcaption[fig3.ps]{The cumulative probability distribution of peak count
rate in terms of logarithm of the peak count rate. 
Open triangles are the observed distribution and the solid lines are the 
best fit curves in the Kolmogorov-Smirnov (K-S) test. 
\label{fig3}}

\figcaption[fig4.ps]{The cumulative redshift distribution of bursts 
with redshifts higher than $z^{'}$ as
a function of redshift $z^{'}$. The upper and the lower panels result from
 the varying $\gamma$ and the varying electron density as well, respectively.  
The solid lines correspond to $A=4$, the long dashed lines to $A=2$,
the dot-dashed lines to $A=1$, the dotted lines to $A=1/2$, and the short 
dashed lines to $A=1/8$.
\label{fig4}}

\clearpage

\begin{table}
\begin{center}
\caption{The maximum redshifts of GRBs which allow the best-fit of models to 
observed data, and the averaged redshifts. 
The parameter $A$ determines the sharpness of the 
window function. Parameters adopted in the calculation are $\alpha=1$, 
$\gamma_{0}=100$, and $\Delta \theta= 1^{\circ}.0$. \label{tbl-1}} 
\vspace{2mm}
\begin{tabular}{ccccc}
\tableline\tableline
&&A&$z_{max}$&$<z>$ \\ \tableline \tableline
constant  $\gamma_{0}$&&N/A&1.60&0.51 \\ \tableline
&&4&3.70&1.03 \\
&&2&3.17&0.91 \\
varying $\gamma$&&1&2.41&0.73\\
&&1/2&1.96&0.61\\ 
&&1/8&1.70&0.54\\ \tableline
&&4&11.23&2.36\\
&&2&3.47&0.98\\
varying $\gamma$  and $N_e$&&1&2.20&0.67\\
&&1/2&1.85&0.58\\
&&1/8&1.60&0.51\\ \tableline \tableline
\end{tabular}

\end{center}
\end{table}

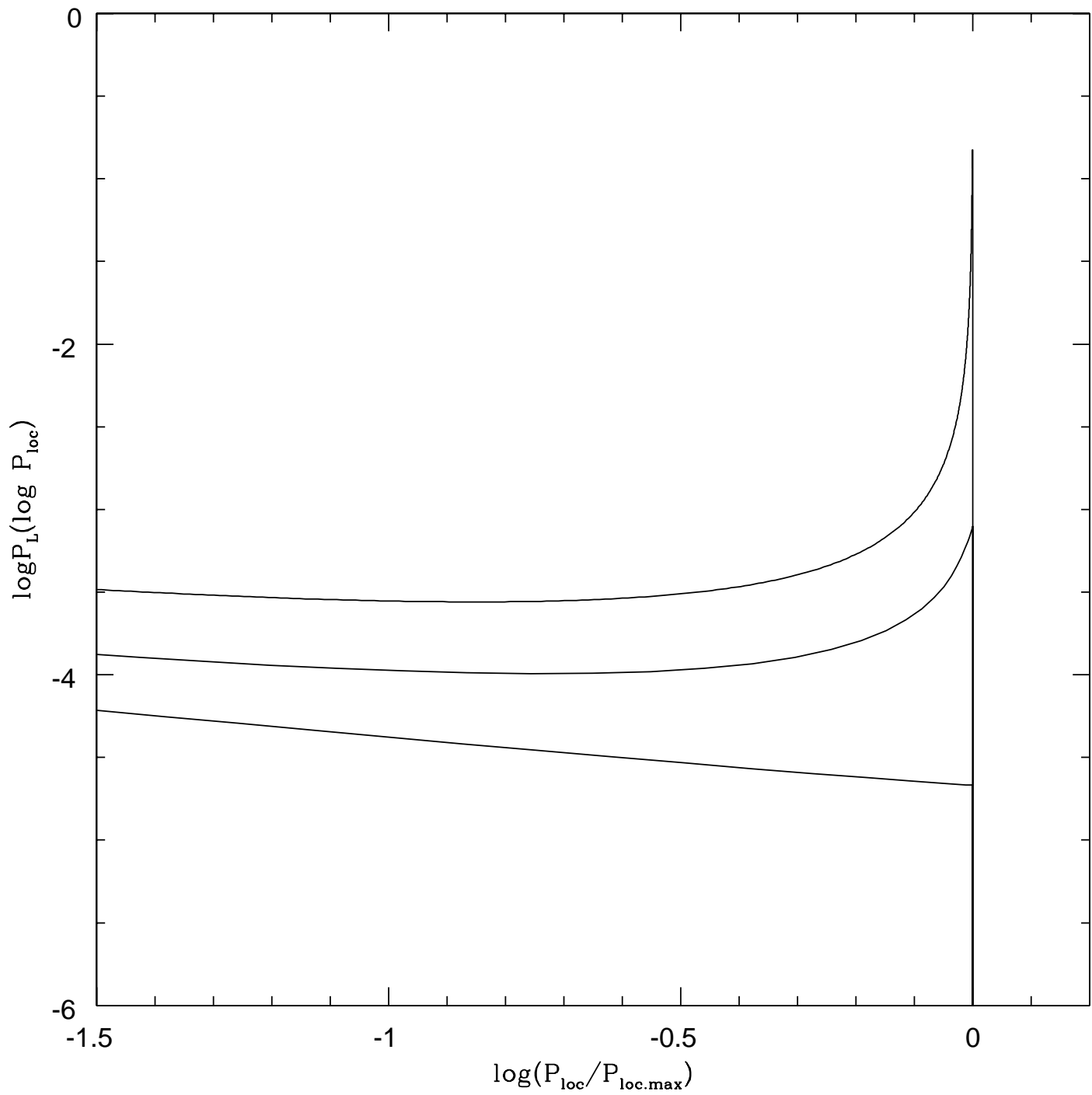
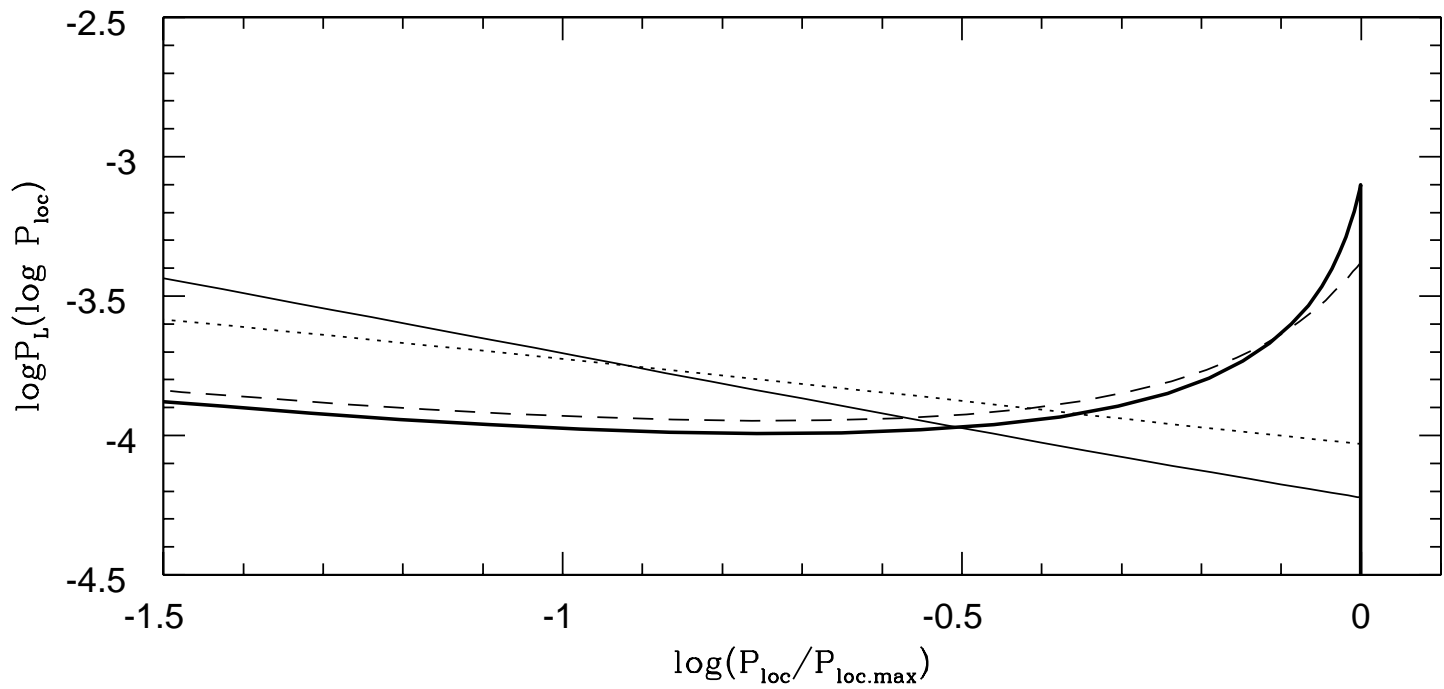
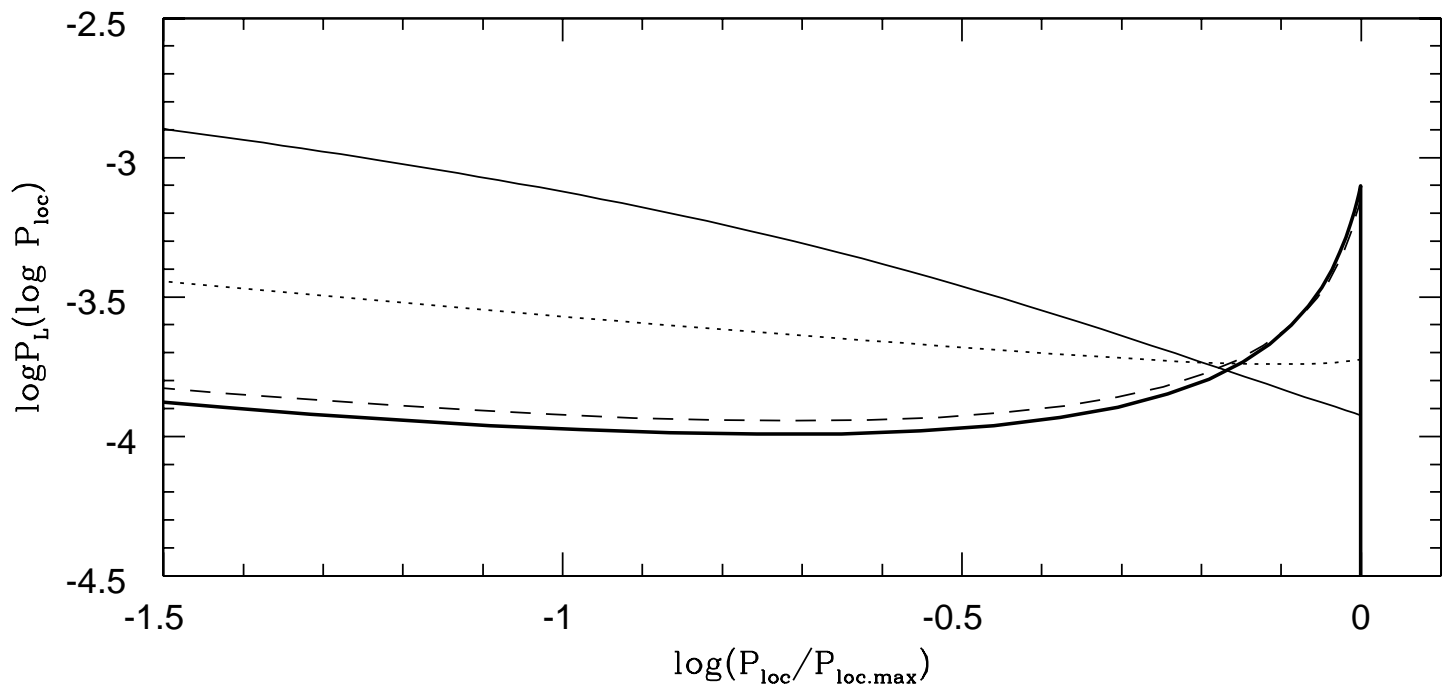
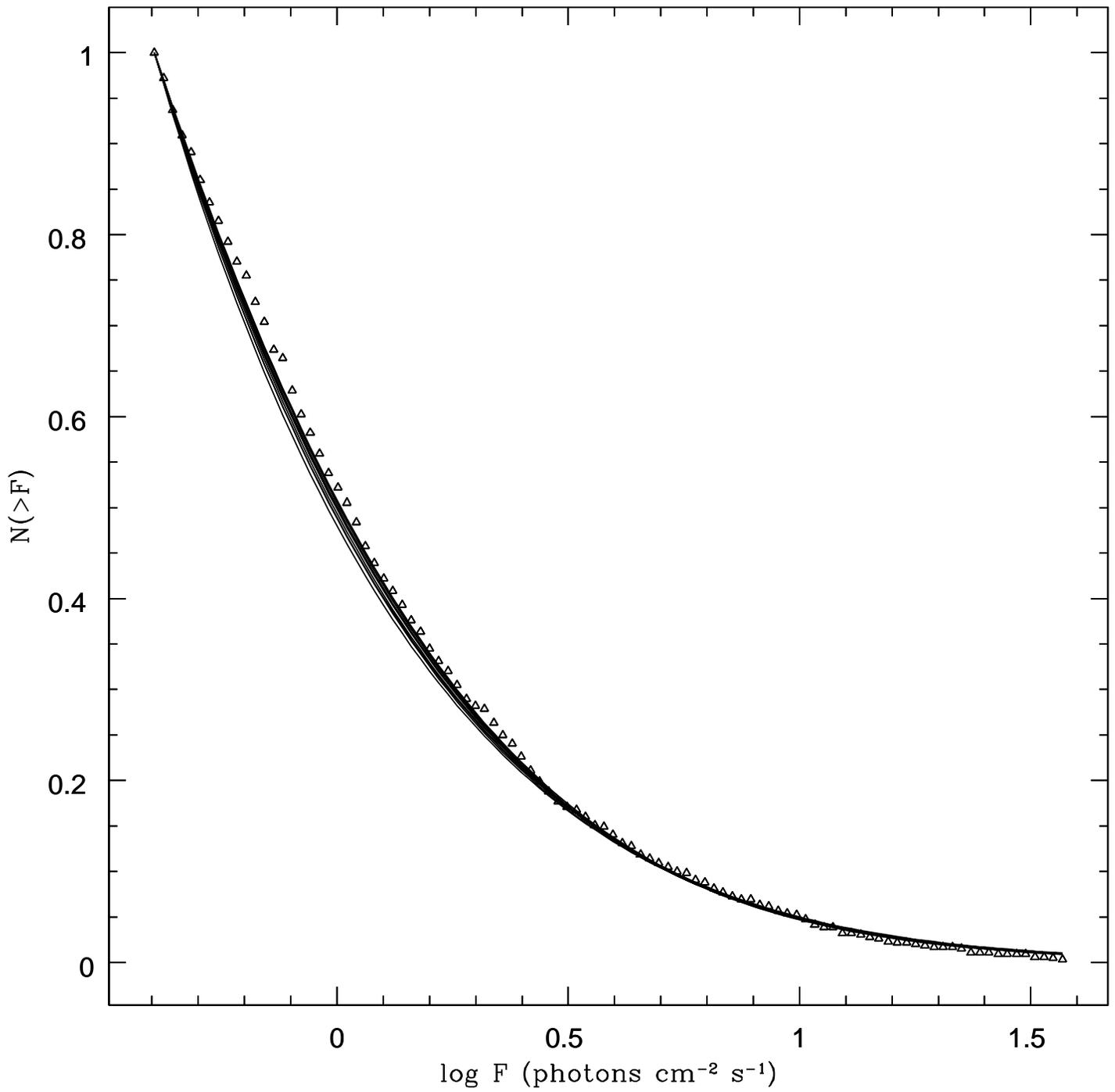
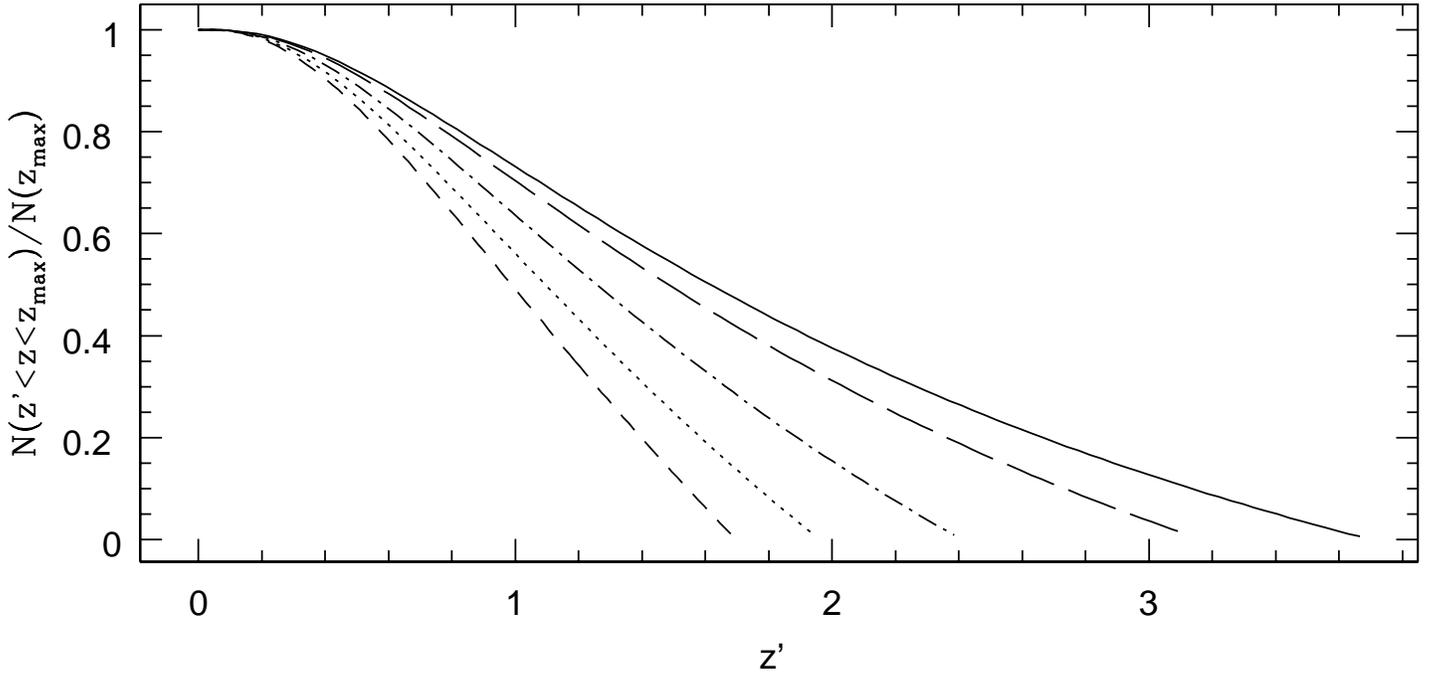
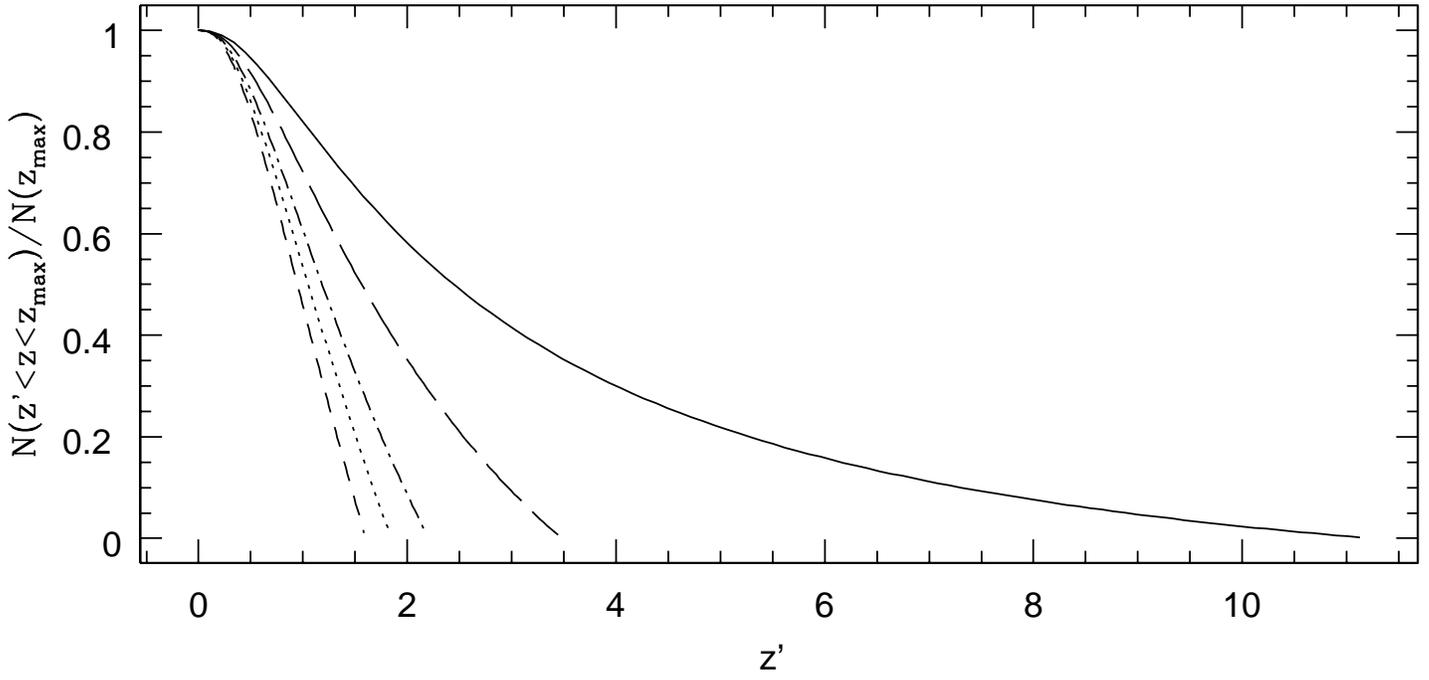
\end{document}